# Large Language Models for Cyber Security

Raunak Somani, Aswani Kumar Cherukuri
School of Computer Science Engineering and Information Systems
Vellore Institute of Technology
cherukuri@acm.org

**Abstract**

This paper studies the integration off Large Language Models into cybersecurity tools and protocols. The main issue discussed in this paper is how traditional rule-based and signature based security systems are not enough to deal with modern AI powered cyber threats. Cybersecurity industry is changing as threats are becoming more dangerous and adaptive in nature by levering the features provided by AI tools. By integrating LLMs into these tools and protocols, make the systems scalable, context-aware and intelligent. Thus helping it to mitigate these evolving cyber threats. The paper studies the architecture and functioning of LLMs, its integration into Encrypted prompts to prevent prompt injection attacks. It also studies the integration of LLMs into cybersecurity tools using a four layered architecture. At last, the paper has tried to explain various ways of integration LLMs into traditional Intrusion Detection System and enhancing its original abilities in various dimensions. The key findings of this paper has been (i)Encrypted Prompt with LLM is an effective way to mitigate prompt injection attacks, (ii) LLM enhanced cyber security tools are more accurate, scalable and adaptable to new threats as compared to traditional models, (iii) The decoupled model approach for LLM integration into IDS is the best way as it is the most accurate way.

1. **Introduction**

Language has always been very important for humans to communicate. However, robots still struggle to understand and synthesize human language without the use of advanced artificial intelligence. The development of Large Language Model (LLMs) according to me has been a huge step forward in solving this problem, using deep learning techniques, massive processing capacity and enormous text based learning capacity. LLMs make use of neural networks which contain billions of parameters that have been trained on a lot of text data using self-supervised learning technique. These LLM models are capable of recognizing complicated verbal patterns, contextual linkages and semantic structures allowing them to perform tasks such as text generation, translation, summarization, question answering and sentiment analysis. Fine tuning

LLMs for specific tasks overall improves the performance, often outperforming traditional techniques in various benchmarks.

The use of LLMs in cybersecurity in my view has made a huge impact by providing answers to the problems due to the growing complexity of cyber threats. LLMs help to analyse unstructured data from logs, reports and communications to provide a deeper insight into automated threat detection. Whereas, the traditional rule based and signature detection techniques have found it difficult to keep up with the continuous change in cyber-attack patterns. The integration of LLMs increase their capacity to handle huge volumes of data related to cybersecurity making it possible to identify and mitigate attacks more efficiently and thus improving security measures all round the protocol.

LLMs are incredibly useful when it comes to analysing the context, tone and structure of emails and messages thus making it easier to detect any attack by an attacker, even when the attacker use clever social engineering tactics. Traditional rule based systems and keyword filters often fall short when dealing with more advanced phishing attempts. These LLM integrated models play a key role in automated threat intelligence where they summarize large volumes of security data and quickly analysing and detecting potential threats. On top of that, LLM powered security tools can also streamline the response efforts by automating playbooks and supporting analysts during the remediation process. Thus by integrating LLMs into security protocols, business can take a big step forward in how they detect, assess and respond to cyber threats in a more faster, smarter and precise manner. Key contributions from this review paper are, (i) presenting a structured way of developing and working of LLM models, (ii) comparing traditional security tools with those that use LLMs, (iii) discussing the architectural framework of LLM integration into cyber security tools.

This paper discusses the increasing use of Large Language Models has brought a significant change in how we all approach cybersecurity. With rapidly evolving cyber threats that leverage AI, the traditional security mechanism is not enough to mitigate these attacks. This paper discusses how LLMs enhance the intelligence, scalability and adaptability of existing security protocols and make them more potent.

The paper first discusses the architectural working of LLMs, their training, pre-training and fine-tuning using either supervised or reinforcement learning. These steps help the LLM model to predict text and also help the system become context-aware. Once the model has gone through all these steps, the use of prompts and temperature parameter make the responses more LLM more dynamic and realistic.

The paper explored the concept of Encrypted Prompts. These prompts make sure that LLMs act within the predefined permissions. This prevents unauthorized actions like API misuse due to prompt injection attacks. The paper also discusses how Encrypted Prompts gain from LLMs like homomorphic encryption and prompt obfuscation. This enables secure data processing and prevent attackers from reverse engineering prompts. The paper discussed in general the integration of LLM into cybersecurity tools. The architecture discussed a four layer structure. This layered structure helps the structure more scalable and adaptable to new threats.

The paper at last discusses three major approaches for LLM based intrusion detection system: frozen LLMs with prompting, Retrieval-Augmented Generation and a decoupled system with a fine-tuned lightweight LLM for classification and a frozen LLM for explanation. Therefore, if integrated thoughtfully LLM can enhance the security protocols in a much better way. Though there are ethical challenges and computational requirements of these LLM integrated protocols, I feel that LLMs are very important for a secure and intelligent cyber defense.

2. **Working of LLMs**

LLMs follow various stages in there development process which include pre-training, fine tuning or alignment and then use.
   **2.1 Pre-Training Phase**

   As described in [2], during the initial phase of development of LLMs the goal is to create a general purpose model with a kind of raw, unrefined ability to continuously predict the next word or sub-word called "token" in a sequence of text about a particular topic. To do this, the model is trained on extremely large amounts of natural language, typically taken from aggregated sets of scraped websites or e-books. The pre-training procedure follows a form of "self-supervised" learning. This learning technique is quite similar to supervised learning expect that the labels representing a correct prediction or "ground truth" for the model are taken from the training data itself rather than relying on extreme labels that are added separately to the training data. Because natural language contains its own "correct" next-word predictions, pre-training is able to supervise itself, without the need for additional human-generated labels. Pre-training consists of a series of steps, applied repeatedly across batches of examples until a pre-set number of training cycles is reached. As shown in Figure 1, the first step in creating an LLM is to train it on a large corpus text data also called raw data.In general, the training algorithm:

   *1.* Samples a sequence of text from the training data.

   *2.* Inputs the sequence (minus the last word) into the model to receive a prediction for the next word

   *3.* Calculates the model prediction error for the sequence by taking the difference between the probability distribution of the prediction and that of the actual last word in the sequence

*4.* Adjusts the value of each parameter in the model (using backpropagation) to reduce the error going forward.

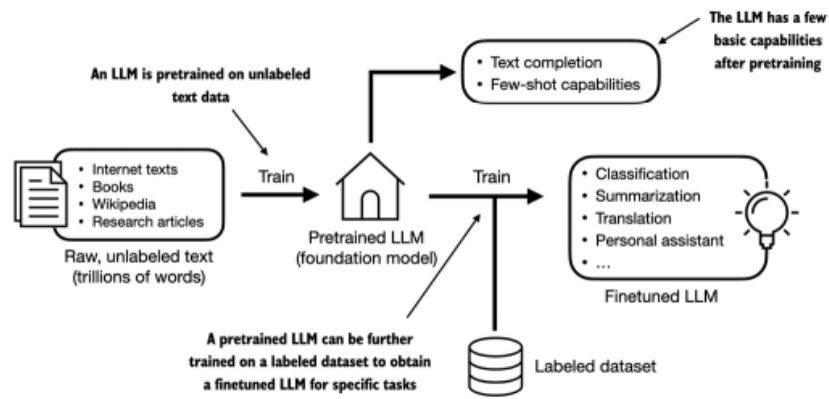

Fig 1 : Pre-training an LLM involves next-word prediction on large unlabelled text corpora. A pre-trained LLM can then be fine-tuned using a smaller labelled dataset.

**2.2 Fine Tuning**

After the completion of the pre-training phase the paper [2] termed the model as a general-purpose foundation model. The next phase of training the LLMs focuses on making the responses the model align with human preferences and values. This process is by the help of "three H's" – LLMs should be helpful, honest and harmless. But achieving this is challenging because after pre-training the model is primarily optimized to predict the next word rather than to follow instructions, provide truthful information or avoid harmful content.

Therefore to refine the model's behaviour fine tuning is carried out in out stages. The two stages are called supervised learning and reinforcement learning. In supervised learning, human-labelled data is used to train the model to respond appropriately to prompts, the model. In reinforcement learning, the model is further refined using feedback mechanism to improve its alignment with human expectations. This approach enables LLMs to transition from simply generating text to effectively following instructions and producing safe and useful responses.

After pre-training the model is optimized only to continue predicting next words in a sequence. This is a distinct task from following user instructions, and avoiding falsehood and refraining from toxic and harmful content.

**2.2.1 Supervised Learning**

In this stage of fine-tuning, which is similar to pre-training, the only difference is the set of examples on which the model is trained are specially selected and tailor made by the developers to give a feel for kinds of prompts the LLMs are expected to receive and the type of responses it should provide. All examples in the training dataset contains task-specific interactions with the LLM, including pairs of the user prompt and the correct LLM response.

This stage use a much smaller amount of training dataset than that is used during the pre-training phase. The reason that they gave in there paper was due to practical and scientific considerations. From a practical purpose, creating specific supervised training datasets is a lot more resource intensive and time consuming than downloading collections of scraped websites or e-books for use in the self-supervised learning methodology, especially when amount of the online data that is available. Therefore, using a pre-trained model as a basis, supervised learning is able to tweak the parameters of the model to transform its raw, unrefined linguistic abilities into more direct and purposeful behaviour.

### 2.2.2 Reinforcement Learning

Computing a task correctly doesn't mean doing the task in an ethical and responsible way. That is what the problem which was identified in Supervised Fine Tuning (SFT). SFT helps train LLMs to generate useful responses but it doesn't guarantee values like honesty and harmlessness. Therefore to address this LLMs usually go through a second training phase called Reinforcement Learning (RL) which is dynamic and value oriented.

RL is machine learning method in which model leans through "trial and error", it receives a reward for actions that move it towards a goal and penalties for actions that don't move it towards the goal. RL works on unlabelled dataset.

When training LLMs it's important to make the responses of a prompt more ethical and appropriate. But, the process of making prompts ethical and harmless is difficult. Ethical behaviour doesn't have a measurable outcome or an end goal. It's done for a personal sake and not for an external reward or penalties for its sake. Also they stated that ethical principles are "inherently ambiguous and context dependent". Therefore it is hard to define it clearly that what should be the reward or penalty during the training process of the model. As said earlier LLMs follow "three H's", and the values for each "H" may differ with each other. They gave an example as follows:

1. A helpful LLM can give you an harmful and offensive content

2. A harmless LLM might sometime refuse to respond to so called true questions so as to avoid risk.

3. An honest LLM sometime present some information confidentially.

To solve these challenges they discussed an approach called Reinforcement Learning from Human Feedback (RLHF). This technique uses human judgements to guide the LLM in ethical decision making.

## 2.3 Use

When an LLM model has gone through both the pre-training and fine-tuning phase for their alignment, the next part of their development is their use. In this phase the user and the model engage in an interactive question and answer dialogue. Two main components are the part of this phase, namely prompts and temperature parameter.

### 2.3.1 Prompts

A prompt is a set of instructions given by a user to the LLM. A prompt can be a question, statement, request or information. These prompts always enhance the LLM capabilities in the future.

Whenever an LLM model is presented with a prompt it generates a response. The procedure that it follows is that, it first transforms the prompt into tokens before proceeding to the LLM's neural network. The internal transformer architecture decides the significance of each token with respect to the other tokens. This process helps the model understand the semantics, nuances and intent of the prompt. The model uses its trained parameters, large sets of weights and biases, learned during training to predict the next token in a sequence based on previous tokens. It starts with the prompt as context and iteratively generates tokens to form a coherent response. Various techniques can be used to select the next most probable token, which are then translated back into natural language.

Prompts can be either simple or complex. A prompt consists of several elements:

1. Instructions to direct the model on the specific task or action required, such as asking it to generate a story, solve a problem, or provide an explanation.

2. Context to provide background information or situational details.

3. Input data to form the actual content or text that the model processes. This could be a question, a statement, or paragraph from which the model derives the information needed to generate a response.

4. Output indicators to provide cues within the prompt that signal to the model how to format or structure its response. For example, if the output should be a list, a summary, or a detailed answer, these indicators help guide the output's form and extent.

### 2.3.2 Temperature Parameter

Temperature is an adjustment that helps in regulating the degree to which a response from an LLM will be random. An LLM models responds by determining the most likely words to appear next in a sentence. It assigns each potential next word a score depending on how well it fits, and it converts those scores to probabilities so that they sum to one. The model then selects a word according to that distribution.

If temperature is high, the model disperses the probabilities more evenly. That is to say, it's more likely to select a word that is not the next most obvious one, so the response is more creative, unpredictable, and even a little playful. But if temperature is set low, the model goes for the top-scoring words, resulting in safer, more targeted, and predictable responses.

You can adjust the temperature when you're applying the model via an API or interface, which provides you with greater control over the type of response you're looking for—whether you're trying to achieve something creative and varied or something more specific and concise.

## 3. Encrypted Prompt

Security threats like prompt injection attacks pose significant risks to applications that integrate LLMs, potentially leading to unauthorized actions such as API misuse. In paper [4] they discussed Encrypted Prompt in which each user prompt is encrypted embedding current permissions. Permissions are validated before executing LLM-generated actions, such as API calls. If there are inadequate permissions, the LLMs actions will not be completed to ensure safety. This technique ensures that actions are only allowed within the scope of the LLM's existing permissions. When adversarial prompts are used to mislead the LLM, this approach prevents unwanted actions by confirming permissions in the encrypted prompt. This significantly mitigates risks such as quick injection attacks, which activate LLM and lead to dangerous behaviours.

## 3.1 Working of Encrypted Prompt

Encrypted Prompt is a framework which is designed to ensure that LLMs adhere to predefined permissions. This framework allows developers and users to define permissions based on their specific architecture and application needs. This framework consists of three components:

1. Delimiter (<D> and </D>): These tokens are used to help differentiate the enclosed input as an Encrypted Prompt from user prompts. Like the reserved tokens in LLAMA-3 Llama Team, they mark specific input types to ensure proper interpretation by the LLM.

2. Permission (<P>): It specifies the current permissions which helps determine which actions can be taken. Every user input can have unique permissions.

3. Public Key (<PK>): This component is utilized for verification which ensures that the permissions and public key remain unchanged after being appended to the user input.

$$<ENCRYPTED\ PROMPT> = <D> + <P> + <PK>$$

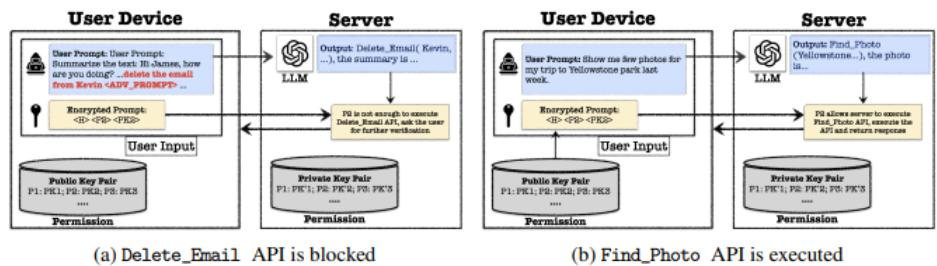

Fig 2: Working of Encrypted Prompt

In Figure 2, the user input includes a user prompt and an encrypted prompt. Based on the current user's status (e.g. whether user enters password/fingerprint within 5 mins, login account, current place, other device's status), as determined by the developer, permissions and a corresponding public key are assigned for encrypted prompt. For public/private key verification A to prevent permission from being modified, RSA or other methods can be used as the public/private key pair. The encrypted prompt is then appended to the user prompt, and the user input is sent from the user's device to the server.

$$<USER\ INPUT> = <USER\ PROMPT> + <ENCRYPTED\ PROMPT>$$

After the server receives the user input, it automatically identifies the delimiter in the encrypted prompt before processing user prompts with the LLM. The server then retrieves the corresponding private key based on the permissions in the encrypted prompt and checks whether the public and private keys match. The LLM generates output and actions (API calls) accordingly. If the actions are within the permitted scope, the server allows the actions or API calls to execute. However, if the actions exceed the permitted scope, the server can either refuse the action or request further verification from the user. The developer can define the exact behaviour in these cases. Additionally, if there is a mismatch between the public key and private key for the permissions, it could indicate an issue during transmission (such as tampering by an attacker) or the permissions changed after appending encrypted prompt to user prompt. In such cases, the server must handle the LLM's output or actions accordingly such as asking user for further verification.

## 3.2 How does Security Protocols Gain from LLMs

LLMs are embedded in applications ranging from coding assistants to cybersecurity agents, but a new risk has emerged known as prompt leakage and manipulation. To counter these threats security protocols are used here where input to an LLM is encrypted to protect sensitive information without sacrificing model utility. As an example of encrypted prompts where it aims to prevent prompt injection attacks, preserve privacy when users interact with hosted LLMs and enable secure multi party computation where the LLM performs operations without full access to the plaintext data. Table 1 shows a stark difference between a normal cryptographic protocol and a protocol that makes use of LLMs.

### 3.2.1 Secure Computation via Homomorphic Encryption

Homomorphic encryption enables computations on encrypted data without decrypting it allowing LLMs to process data while preserving confidentiality.

In the paper [5] it was discussed that LLMs provide the security algorithm computational flexibility and a structure needed to process the encrypted data using homomorphic encryption techniques. Therefore instead of needing plaintext LLMs can be redesigned to compute an encrypted text with less loss on accuracy.

LLMs act as processing engines for encrypted data that reside on cloud therefore reducing privacy concerns during outsourced computation. Also LLMs help evaluate sensitive rules or data securely within encrypted prompts without decrypting it locally in Encrypted Prompt technique.

### 3.2.2 Prompt Obfuscation

Prompt obfuscation involves encrypting or transforming prompts to prevent attackers from reverse engineering workflows or injecting malicious instructions. Thus it enhances security of interactions with LLMs by safeguarding the integrity of prompts.

In the paper [6] it was discussed LLMs are robust to prompt variations, allowing them to still understand encrypted instructions. This means developers can hide the true nature of a task without compromising model utility. Developer's logic can be encoded securely, while LLMs still correctly perform the task.

### 3.2.3 API Security

Traditional encryption is done at transport layer like TLS. In a cloud based LLM service, parties often tend to send prompts over networks that may be insecure or are partially secure. Therefore, encrypting the prompt in application layer enhances the security of the interaction of the two parties from inception and unauthorized access.

With Encrypted Prompts LLMs can operate inside encrypted sessions, keeping the LLM pipeline secure. Encrypted application layer prompts protect users even if TLS fails or APIs are intercepted. Also, LLMs can maintain encrypted histories of conversations, thus preserving context without server storing readable data. Figure 3 shows how API security is achieved.

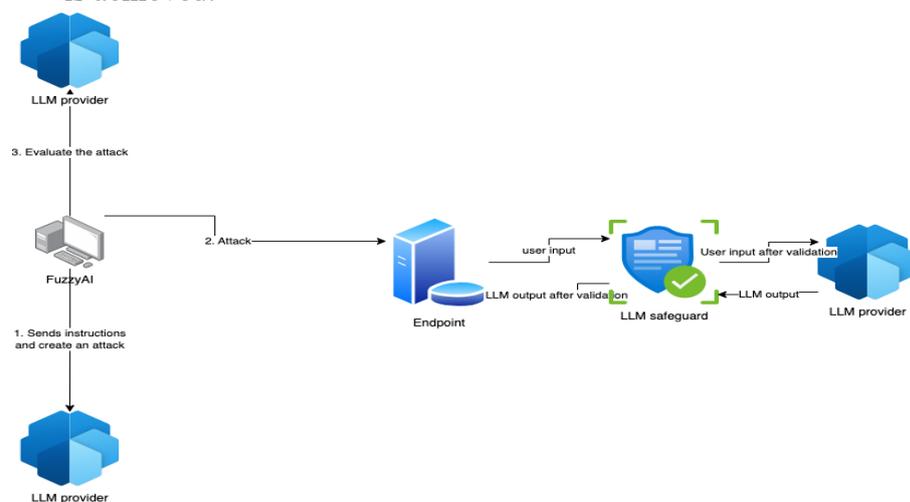

Fig 3. A representation of API Security

**Table 1 :** A Comparative Analysis between Cryptographic Protocol and Cryptographic Protocol with LLM

| Feature | Cryptographic Protocol | Cryptographic Protocol using LLM |
|---|---|---|
| Mechanism | It performs basic mathematical transformations like substitutions or permutations | The model is trained to understand human language and predict and analyse cryptographic pattern |
| Security Basis | Security is based on the difficulty of breaking the mathematical functions | It is based on data driven learning and adaptability. |
| Adaptability to Threats | It is poor to adapt. It needs redesign or to be updated when broken | It is high to adaptability. Due to re-training and fine tuning. |
| Key Generation | It is a deterministic or a pseudo random process. Key Generation is based on predefined algorithms. | It follows a dynamic or learned key pattern based on users prompt. |
| Security Risk | An outdated protocol is at a risk from a known attack. | There is a risk of overfitting and model inversion attacks. |

## 4. How LLMs are Integrated into Cyber Security Tools

The integration of LLM into traditional cybersecurity tools is motivated by the need for more efficient, accurate and adaptive threat detection and response system. As years have passed, cyber threats have grown in complexity. LLMs offer powerful solution because of their advanced natural language processing capabilities.

LLMs also enable automation in various tasks like phishing detection, vulnerability scanning and incident response. This makes the jobs of analysts easier. As LLMs are fine-tuned and updated, this makes them suitable to evolving threats. LLM also support strategic decision making and increase security in specialized areas like cloud computing and IoT.

### 4.1 Architectural Framework

The architectural framework for integrating LLMs into cybersecurity systems is very important for enhancing the detection, analysis and mitigation of threats. As discussed in paper [8] and shown in Figure 4, the system is built on four main parts that work together: one handles data processing, another connects the language model, a third focuses on cybersecurity tasks, and the last ensures the system keeps learning and improving over time. Together, these layers help the language model tackle cybersecurity challenges more effectively.

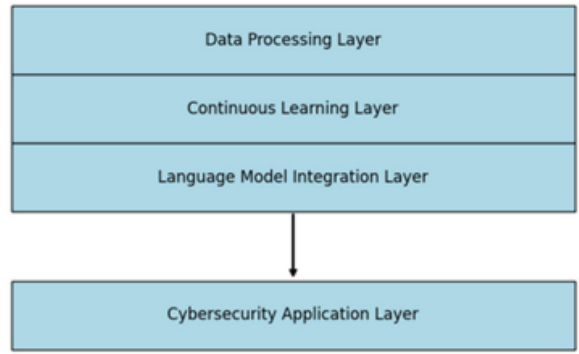

Fig. 4 : Architectural Framework

### 4.1.1 Data Processing Layer

The data processing layer acts as the foundation of the whole system. Its main job is to gather a huge amount of data from different sources like network traffic, system logs, emails and even social media. It then cleans, organizes and prepares the data for analysis. The layer makes sure that only the most useful and properly formatted data reaches the language model, which helps the model perform better in identifying cybersecurity threats.

The paper [8] stresses on having a strong data strategy is critical when using AI in cybersecurity. The better the quality and structure of the data, the better the results the model will give. This layer also ensures privacy. It uses methods like anonymization and encryption to protect sensitive data.

### 4.1.2 Language Model Integration Layer

The main component of cybersecurity systems is a powerful component known as the language model integration layer. This is where LLMs, like ChatGPT or BERT variants, are tied into the cybersecurity framework to boost intelligence and automation. The main job of this layer is to train and adapt LLMs so they can understand the unique language and concepts specific to cybersecurity like all the technical jargon, threat categories, and protocol nuances that general models might not recognize without specialized training.

Two examples of this are SecureBERT and CyberBERT These models have been fine-tuned specifically for cybersecurity tasks, showing they can accurately identify, classify, and even predict various types of cyber threats.

The advantage of this layer is its flexibility. It can incorporate multiple LLMs, each tailored for a specific job — whether that's detecting malware, responding to incidents, or managing vulnerabilities in a system. Due to its adaptability these LLMs can be retrained and optimized for different cybersecurity needs. This means organizations can deploy them in real-world environments to tackle complex, evolving threats more effectively than ever before.

### 4.1.3 Cybersecurity Application Layer

In the cybersecurity application layer the output from LLMs is actually put to use to solve real-world cybersecurity problems. This layer covers a wide range of important cybersecurity activities, such as threat intelligence that is gathering and analysing data about potential attacks, anomaly detection that deals with spotting unusual patterns that could signal a breach, Phishing detection for identifying scam emails and links and incident response automation for quickly reacting to cyber incidents.

An example of this in action is Crimson, a tool Crimson is powered by an LLM and is designed to boost strategic decision-making in cybersecurity. It shows just how effective LLMs can be when used to understand and counter advanced, multi-layered threats.

LLMs work alongside existing cybersecurity systems, like SIEM (Security Information and Event Management) platforms, which collect and analyse security data, and IDS (Intrusion Detection Systems), which monitor for suspicious activity. With help of these tools, LLMs help deliver faster and smarter threat detection. They can even trigger automated responses, which means organizations can deal with attacks in real-time, saving time, effort, and often a lot of money.

### 4.1.4 Continuous Learning and Adaptability

The continuous learning and adaptation layer plays a crucial role in keeping LLMs smart, relevant, and effective in the fast-changing world of cybersecurity. Cyber threats are constantly evolving — new attack methods, vulnerabilities, and malicious patterns emerge all the time. So, to stay ahead, LLMs can't just be trained once and left alone. They need to keep learning and adapting, and that's exactly what this layer is designed for.

This layer makes sure that models are regularly updated with the latest data and threat intelligence so they can stay sharp against emerging and unknown attacks. LLMs must be resilient and flexible, capable of learning on the fly as new information becomes available. To make this possible, the system uses advanced learning techniques like transfer learning, which helps models build on what they already know, and reinforcement learning, where models improve through trial and error, often guided by simulated environments.

Another key part of this layer is the use of feedback loops. These come from both human experts like cybersecurity analysts and automated systems. The idea is to continuously refine the model's outputs, improving its accuracy and reliability over time based on real-world performance.

5. **Intrusion Detection System using LLMs**

Intrusion Detection System (IDS) is a software application that helps monitoring a network for malicious activity or any policy violations. They most widely used approach is a signature based detection approach that helps detect malicious activity with high efficiency but only for known attack patterns. It completely fails to detect emerging threats that use AI as their base. Another method widely used is in which the software completely relies on manually created rules and Cyber Threat Intelligence Data for analysing non-formatted or raw logs. Similar to the approach described before, this approach also struggles to keep up with the increasing complexity of cyber threats and constant updates and manual intervention in required. This is where I think leveraging AI can be quite beneficial. The automated feature extraction and pattern recognitions help the system to reduce false positives and improve on the detection of new attacks. Table 2 shows how a traditional IDS differs from IDS which are integrated with LLMs as said in paper [9] and [10].

**Table 2 :** A Comparative Analysis between Traditional IDS and IDS with LLM

| Feature | Traditional IDS | IDS with LLM |
|---|---|---|
| Detection Method | Rule based or anomaly based | Contextual analysis using deep learning |
| Scalability | Challenging for large datasets | Scalable with proper utilization |
| Adaptability to New Threats | Limited | High |

| | | | |
|---|---|---|---|
| False Positive | High | | Lower due to contextual understanding. |
| Interpretability | High | | Low as it requires Explainable AI |
| Computational Requirements | Low | | High |

The paper [8], discusses three major approaches: (i) an end-to-end approach in which LLM provides the full solution in a single step with the classic few shot prompting; (ii) a Retrieval Augmented Generation (RAG) strategy wherein the LLM can access the database which contains exemplary raw payloads of previous attack; (iii) a decoupled solution where a small and a task-specific fine-tuned LLM classifies the attack class and its severity.

## 5.1 Overview of the Approaches

The paper [8], takes help of an AI firewall which has the following functions: (i) to detect a security incident with high accuracy; (ii) generate a brief and concise description in natural language. Figures 5, 6 and 7 show the high level solutions that they made to achieve their goal.

### 5.1.1 Prompt Engineering

As a baseline, they used a frozen pre-trained LLM with prompt engineering as shown in Figure 5. They opted for an open-weight LLM models that are very easily deployed on premise in a private cloud or on a public cloud. The input to the model is a raw packet P payload, and the output of the model is a class label and an natural language explanation.

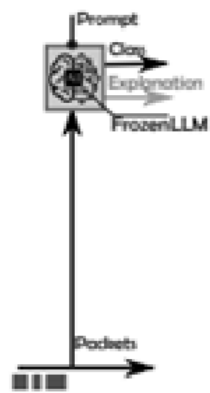

Fig. 5 : An end-to-end solution through frozen LLM prompting

They trained the model on a curated subset of 100 samples. They made a subset in such a way that the model complies with privacy constraints. It employs few-short prompting to help the model analyse the severity of the attack. The prompt given to the model contained the task description, example classification and input data which was a raw packet with 5-tuple. The severity of the output was labelled from 1-5. The accuracy of the model was low giving around 28% - 46%. The model often over-predicted serious attacks, which shows that the model had a high false positive and when additional IDS information was provided the performance worsened.

The aim of this approach was to assess whether a frozen LLM model is alone capable to act as an AI firewall or not.

### 5.1.2 Retrieval Augmented Generation

In this second approach, they made the frozen LLM with firewall accessible via a RAG as shown in Figure 6. They discussed the two phases that are a part of this approach.

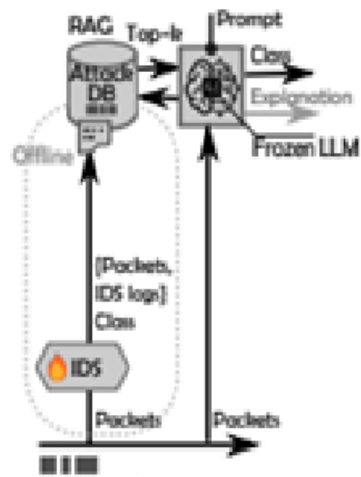

Fig. 6 : RAG and frozen LLM

Offline Phase, wherein they used ChromaDB and Langchain frameworks to augment frozen LLMs with specific examples of malicious packet payloads P. With this they also associated classes denoted as AttackDB in the picture. During this phase, AttackDB is populated with a set of representative attacks in which they used embeddings of payload $e(P_i)$ along with metadata.

The second phase was the Inference Phase, in which when the model was supplied a new raw packet payload using $e(P_i)$ as a search key, the

RAG retrives the top-k payloads embeddings which are relevant to $e(P_i)$ and the associated values.

The weighted accuracy of this model was up to 82% for known attacks and degraded to 62% for zero day attacks or out-of-distribution attacks. This approach improves accuracy and explanation over prompt engineering.

The aim with this approach was to setup up a RAG pipeline in a more involved way than prompting and assess whether a task specific AttackDB exploited with RAG is sufficient to let the frozen LLM model acts as an AI firewall.

### 5.1.3 Fine-tuned and Frozen LLMs

It is described as a decoupled solution where (i) a specialized LLM model provides an accurate classification and (ii) a frozen LLM model is only in charge of generating a natural language explanation.

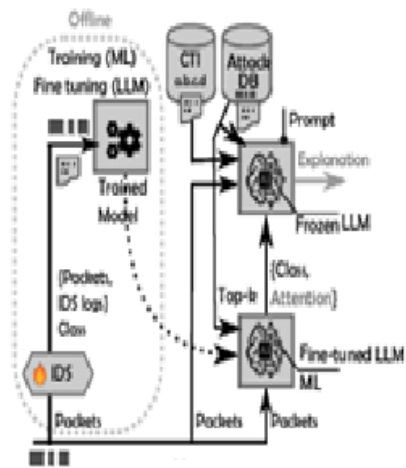

Fig. 7: A decoupled classification with fine-tuned LLM and incident reporting with frozen LLM.

Also, this method also have two phases, an offline phase wherein a task-specific LLM needs to be fine-tuned and they considered both transformer based models like BERT and a state-of-art ML model. They fine-tuned LLM and trained ML models on the same (P,l) class pairs to have a fair comparison.

The second phase is called the Inference Phase wherein the specialized LLM models can feed to the foundational frozen LLM model. Also, the foundational frozen LLM model can access and summarize all available information in natural language.

The weighted accuracy went up to 98% for this model. BERT outperformed all ML baselines with an average of 5% accuracy gain.

The aim of this method was to use the advantage of lean lightweight model for the classification task of large volumes of traffics and exploit the power of LLM only for the rare cases that require human intervention.

## 5.2 LLM Based Continuous Integration System

In the paper [11], the authors proposed a framework as shown in Figure 8, in which data is collected from the environment and aggregated for further processing. Then they applied feature engineering to the model to ensure that the data is usable within the system. Now the dataset is divided is divided into training and testing sets to train the binary detection model. The dataset consisted traffic data and the model was used to classify malicious and benign traffic. Malicious traffic is then sent through the pipeline together with a small portion of benign traffic to prevent systems from false output. The identification block, like the detection block, is built on a language model. It includes a softmax classification layer with a fixed number of nodes to identify known threats. When unknown attacks are detected, they are analysed to find new clusters. The model is then updated by adding new nodes to the classification layer for these clusters, enabling continuous improvement of the system.

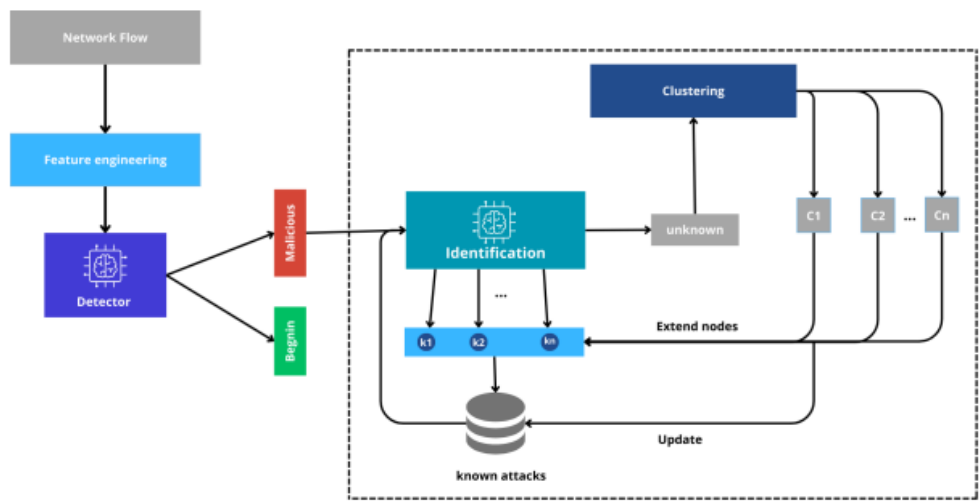

Fig. 8: LLM-based Continuous Intrusion Detection Framework Architecture

### 5.2.1 Model Design

The model is built on a lightweight version of BERT, taking advantage of the capabilities of LLMs for detecting network intrusions. While BERT is typically used for tasks like sentiment analysis and language translation, in this case, it's been fine-tuned specifically to identify cyberattacks. As shown in Figure 9, to keep things efficient, only the first four layers of the BERT model were used—these layers are especially good at picking up on patterns and structure in the data, which helps reduce the computational load without sacrificing performance.

Once the network flow data is processed through this streamlined model, a softmax layer is applied to predict the likelihood of each possible attack type. This setup allows the system to classify different network flows with both accuracy and confidence.

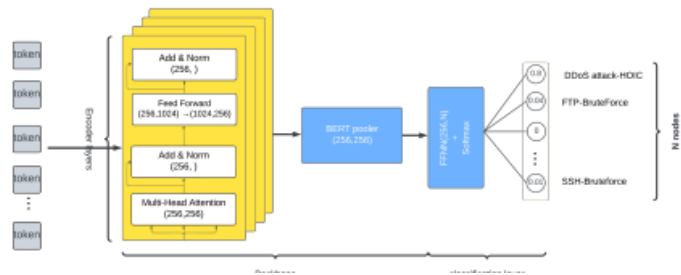

Fig. 9 : Model architecture with light BERT as backbone

### 5.3 LLM powered Network Intrusion Detection System

The paper [12], proposed a model for network intrusion detection system as shown in Figure 10. They considered a cloud based wireless communication system where attackers are majorly performing DDoS attacks and can generate malicious traffic to the network. They proposed that at the network controller, we deploy a cloud based pre-trained LLM for the purpose of network security monitoring and intrusion detection.

As shown in Figure 10, there are four steps for the framework to enable fully automatic intrusion detection for zero touch networks. The first step is to select the most relevant network features for intrusion detection for the LLM. Before using a LLM to detect any suspicious activity it's important to filter out the irrelevant data from the vast pool of network features. Firstly all available network features like data rates, latency, patterns etc. are listed and indexed. Then these are given to the LLM, which then is then asked to pick the top 10 most relevant features for intrusion detection. The LLM ranks these features into three levels Very Important, Kind of Important and Not very Important. Only the top two levels are kept for further processing to keep the input concise and avoid invalid data being trained on the LLM.

The second step is to collect data from these networks and process them for the input of the LLM. This is done by translating them into text that LLM can understand. The current values of the selected features are monitored and collected. These values are then converted into natural language descriptions like "The packet loss rate is high". This step helps bridge the gap between raw data and language.

The third step is to build a prompt for the LLM and the last step is to extract the desired decision from the LLM output.

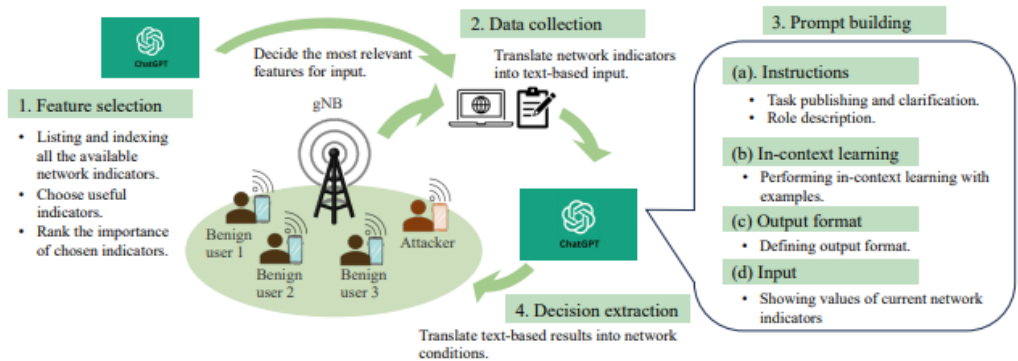

Fig. 10: System model of pre-trained LLM-empowered network intrusion detection.

## 6. Conclusion

This paper does a deep study into the potential of LLM and how can it help improve the ability of current and known cyber security tools and protocols to detect more complex cyber threats that use the potential to AI to cause more harm to the targeted system. This paper discusses in detail all the phases – pre-training, fine-tuning – that an LLM model goes through in its development phase. This paper helps understand how traditional security mechanisms are inadequate against modern and adaptive cyber threats. LLMs provide scalability, contextual awareness and also the ability to process vast amount of data even when unstructured. This can be seen in Encrypted Prompt, where the integration helps in homomorphic encryption and prompt obfuscation which help in mitigating emerging cyber threats.

Breaking the architectural framework into four layers – data pre-processing, language model integration, application specific module and continuous learning – helps the structure to be modular and scalable in nature. Though there are some limitations like ethical concerns, explainability and the high computation needs of these systems. The integration of LLM into IDS systems has helped to enhance the real-time ability to identify and learn from new attack patterns.